# Van der Waals semiconductor InSe plastifies by phase transformation

Yandong Sun[1,±], Yupeng Ma[2,±], Jin-yu Zhang[3,5], Tian-Ran Wei[2*], Xun Shi[2,4],

David Rodney[3*], Ben Xu[1,5*]

[1]Graduate School of China Academy of Engineering Physics, Beijing 100193, China

[2]State Key Laboratory of Metal Matrix Composites, School of Materials Science and Engineering, Shanghai Jiao Tong University, Shanghai 200240, China

[3]Univ. Lyon, CNRS, Université Claude Bernard Lyon 1, Institut Lumière Matière, UMR5306, Villeurbanne 69622, France

[4]State Key Laboratory of High Performance Ceramics and Superfine Microstructure, Shanghai Institute of Ceramics, Chinese Academy of Sciences, Shanghai 200050, China

[5]AI for Science Institute, Beijing 100080, China

[±]These authors contributed equally to the work.

E-mail: tianran_wei@sjtu.edu.cn; david.rodney@univ-lyon1.fr; bxu@gscaep.ac.cn

## Abstract

Inorganic semiconductor materials are integral to various modern technologies, yet their brittleness and limited deformability/processability pose a significant challenge in the development of flexible, wearable, and miniaturized electronics. The recent discovery of room-temperature plasticity in a few inorganic semiconductors offers a promising pathway to address this challenge, but the deformation mechanisms of these materials remain unclear. Here, we investigate the deformation of InSe, a two-dimensional (2D) van der Waals (vdW) semiconductor with substantial plasticity. By developing a machine-learned deep potential, we perform atomistic simulations that capture the deformation features of hexagonal InSe upon out-of-plane compression. Surprisingly, we discover that InSe plastifies through a so-far unrecognized martensitic phase transformation; that is, the layered hexagonal structure is converted to a tetragonal lattice with specific orientation relationship. This observation is corroborated by high-resolution experimental observations and theory. It suggests a change of paradigm, where the design of new plastically-deformable inorganic semiconductors should focus on compositions and structures that favor phase transformations rather than traditional dislocation slip.

Inorganic semiconductor materials play an indispensable role in electronic devices, communication systems, energy technologies, and various other industries that shape modern society[1–3]. However, traditional inorganic semiconductors often display poor deformability and processability, being highly susceptible to failure under external forces or thermal loading during service[4,5]. Due to the rapid development of electronic devices with increasingly complex shapes[6,7], the constraints on material processing have intensified. Additionally, the development of smart wearable devices requires semiconductor materials with high deformability and flexibility[8–10]. Therefore, understanding and enhancing the possible deformability of inorganic semiconductor materials has become an urgent and significant pursuit[11].

Recently, exceptional plasticity in ambient condition has been discovered in a series of bulk inorganic semiconductors, such as $Ag_2S$[12,13] and its alloys[14], InSe-like 2D vdW crystals[15–21], AgCl/AgBr[22] and ZnS[23,24] crystals (in darkness). These findings greatly alter the long-standing preconception of the brittleness of inorganic nonmetallic materials. Meanwhile, these novel materials provide an ideal platform to explore new deformation mechanisms, which will certainly add new facets to the discipline of materials mechanics.

The deformation mechanism is profoundly influenced by the bonding states of materials and the topological characteristics of these bonds. In conventional inorganic semiconductors, directional covalent bonds generate rigid lattices, which exhibit limited deformability only observed at small scales, for example, when straining micro- and nano-samples[25–29] or in the case of confined plasticity generated by nanoindentation[30–34]. In these materials, comparable bond strengths across all three dimensions lead to deformation that reconstructs bonds in multiple directions. In contrast, the majority of 2D vdW materials exhibit bonding that is predominantly layer-confined, resulting in deformation primarily through inter-layer sliding between weakly bonded layers[35]. Yet, the observation in 2D vdW InSe of surface steps and slip traces during compression along the out-of-plane direction (see Fig. 1a) at approximately 45° from the compression axis (Ref. 15), challenges the aforementioned mechanism. Although a cross-layer dislocation slip was proposed to rationalize this phenomenon, the high energy barrier of this slip (5-10 times larger than the interlayer glide, Figs. 1d and e) indicates the difficulty of this process and suggests additional mechanisms for this fascinating behaviour.

To get an insight into the sophisticated deformation of the 2D vdW crystals, we performed atomic-scale simulations of the compression of InSe nanopillars perpendicular to the 2D layers as in the experiments (see Fig. 1b). A challenge is that, while first-principles methods predict accurately the structure and deformation of InSe, such calculations are too computationally-expensive to model nanopillars large enough to allow for inhomogeneous deformation. Fortunately, machine-learned potentials have made strong progress and now allow us to model large complex systems with quantum accuracy. Using the machine-learned deep potential model proposed by Wang et al.[36], we trained an interatomic potential to simulate different phases of InSe and their deformations. We then used the interatomic potential to simulate the uniaxial compression of InSe perpendicular to the van der Waals planes and analyze the deformation mechanisms. Surprisingly, we found out that InSe plastifies through a so far unknown martensitic phase transformation, proposing an alternative to the dislocation slip mechanism proposed earlier[15]. We anticipate that such transformation-induced plasticity might be common to other 2D vdW semiconductors of technological interest.

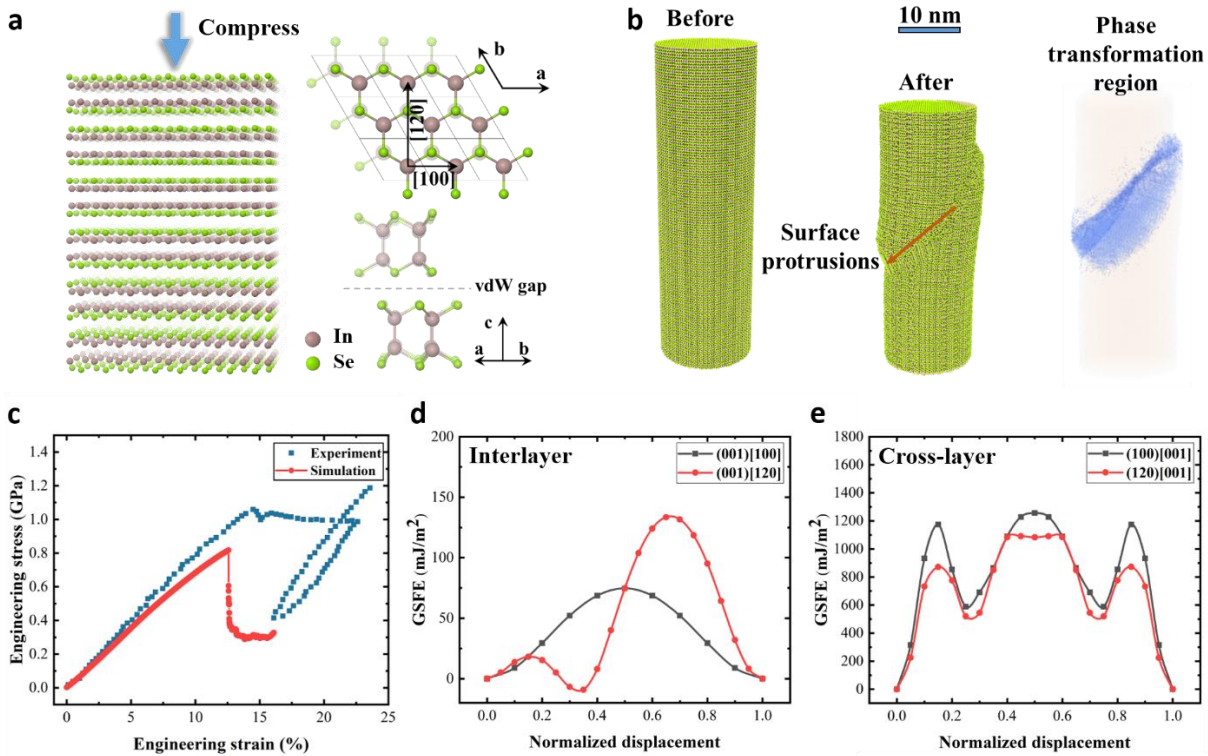

**Fig. 1| Plasticity of InSe under compression**. **a**, Atomic structure of β-InSe with corresponding atomic orientations. **b**, Macroscopic morphological changes in samples before and after compression, highlighting the phase transformation region. **c**, Stress-strain curves for both numerical and experimental compressions (experimental data from Ref. 15). **d**, Generalized stacking fault energy (GSFE) curves for slip in the [100] and [120] directions parallel the 2D layers. **e**, GSFE curves for slip along the [001] direction in two different planes perpendicular to the 2D layers.

## Construction of machine-learned interatomic potential

In this work, a machine-learned deep potential (DP) was developed to accurately and efficiently simulate different phases of InSe, as well as their deformation and transformation processes. The DP approach has proved very efficient in training numerically-efficient deep neural network potentials for various materials[37–39]. The primary training datasets were generated using the DPGEN software[40] and comprised configurations acquired under diverse pressure and temperature conditions, as illustrated in Fig. 2a. To address the challenge that significant and heterogeneous deviations of atomic structures away from equilibrium in deformation simulations, we incorporated into the training dataset various configurations derived from applying a range of deformations to small-scale models. These deformations include stretching, compression, and shear, as illustrated in Fig. 2b. Given the layered nature of InSe, we also included configurations pertaining to interlayer cleavage, showcased in Fig. 2c. Finally, recognizing the importance of dislocations in deformation, data involving generalized stacking fault energy (GSFE), which describe the energy barrier against slip (see Fig. 2d), were also included. First-principles calculations and training procedures are detailed in Supplementary Materials S1 and S2. After training a neural network (schematized in Fig. 2e), we obtained a DP model for the potential energy surface of InSe, conceptually illustrated in Fig. 2f, which can be used for molecular dynamics (MD) simulations. The trained DP model allows to simulate various deformation scenarios of InSe, including homogeneous stretching, compression, and shear included in the training dataset, but also bending, cleavage, interlayer slip, phase transformation, and 2D dislocations, as illustrated in Figs. 2g to j.

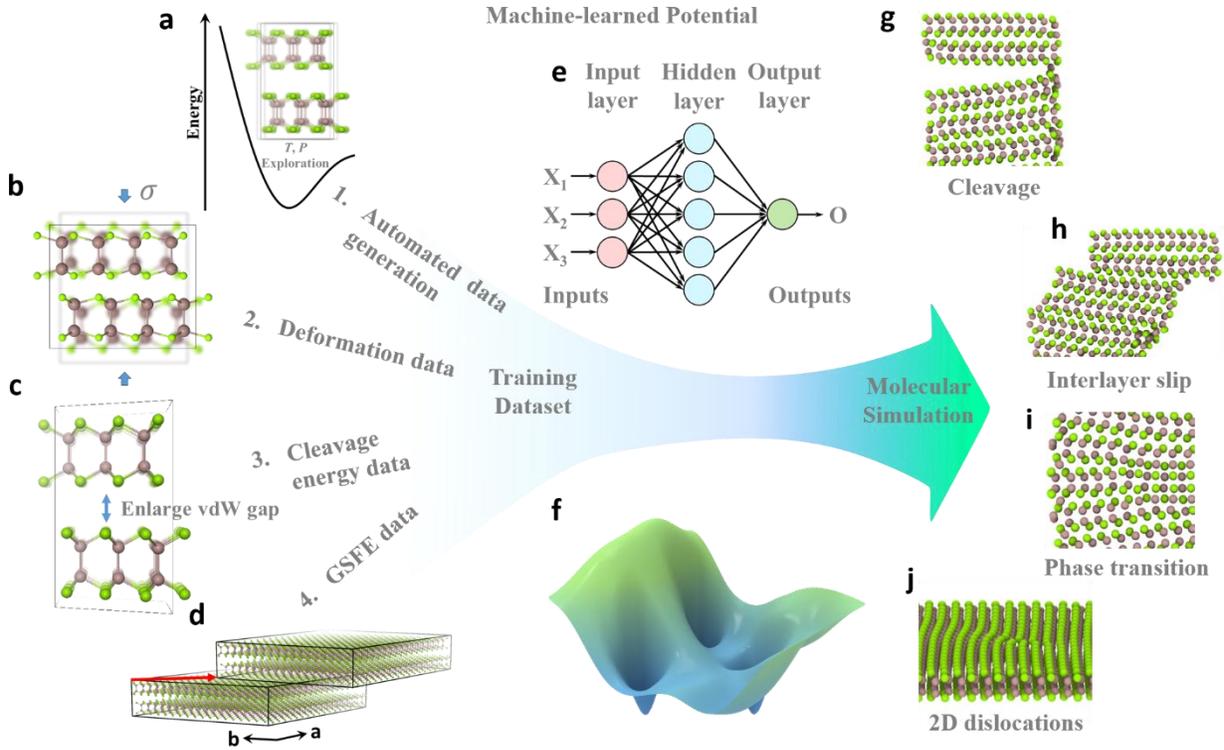

**Fig. 2| Overview of InSe DP potential training process. a-d**, Illustration of the training dataset components: **a**, automated data generation from DPGEN software; **b**, deformation configurations in small-scale models; **c**, cleavage energy data, and **d**, GSFE data. **e,f**, Schematic diagrams of the neural network and potential energy surface, respectively. **g-i**, Various defects involved in deformation.

## Uniaxial compression and stress-strain curves

For both atomistic simulations and micro-pillar compression experiments, the compressive deformation was applied along the [001] direction of hexagonal β-InSe (space group $P6_3/mmc$), as depicted in Fig. 1a. Simulations and experiments reveal a typical elasto-plastic behavior, visible in the stress-strain curves in Fig. 1c, which can be decomposed into three distinct stages.

In Stage I, which corresponds to elastic deformation, the stress increases linearly with strain. This initial process does not result in defects but leads to a reduction in the interlayer spacing due to the contraction of the In-In bonds. When strain exceeds approximately 12.5%, the stress suddenly breaks away from linearity, the onset of Stage II. This is marked in the simulations by a sudden stress drop, not visible in the experiments. This is probably due to the difference in scales, nanometric in the simulations, micrometric in the experiments. At this point, both in the simulations (Fig. 1b) and in the experiments (Ref.15), surface steps form along bands at approximately 45° to the loading axis. Deformation then proceeds at constant stress, again similarly in the simulations and the experiments. This corresponds to Stage III of plastic flow. It is worth noting that the simulated and experimental yield stresses, corresponding strain levels, and deformation features are highly consistent, confirming the accuracy of the present DP model and the reliability of the associated simulation results.

We used the atomic-scale simulations to analyze the crystal structure in the regions of the surface steps. We found out that plastic yield was not due to dislocation slip but to a stress-induced phase transformation, through which the crystal transforms into a tetragonal phase with point group I4/mmm, different from the layered hexagonal structure of the matrix. This tetragonal phase reported in Refs 41 and 42, but not the context of plasticity, and its orientation relationship with the hexagonal InSe matrix are further analyzed below.

## Orientation relationships between new and parent phases

In order to facilitate the study of the orientation relationship (OR) between the tetragonal product phase and the parent hexagonal phase, we constructed a larger orthogonal model with dimensions 16.8 nm in the $x$-direction, 16.2 nm in the $y$-direction, and 54 nm in the $z$-direction, comprising 491,520 atoms. Vacuum layer was introduced in the $x$ and $y$ directions. In contrast to the cylindrical model, the orthogonal model exhibits only two free surfaces, parallel to low-index crystal planes of hexagonal InSe: (100) and (120). During the compression simulations, a significant number of regions of the tetragonal phase were observed (Fig. 3a). These regions demonstrated two specific ORs with the parent phase. Fig. 3b presents a magnified view of an example of a two-phase region. A perfectly coherent interface is visible between the transformed tetragonal phase on the right-hand side and the initial hexagonal phase on the left-hand side. The interface makes an angle of approximately 45° with respect to the layered planes of the hexagonal phase. Detailed analysis of the atomic structure of the tetragonal phase (Fig. 3c) reveals that it possesses the I4/mmm point group symmetry. We calculated the lattice parameters for the tetragonal phase and observed a 16.25% reduction in volume compared to the parent hexagonal phase. This significant volume difference between both phases is at the root of the plastic deformation. We also calculated the phonon spectrum of the tetragonal phase, confirming its stability under a pressure of 3.0 GPa. For more details, please refer to Supplementary Materials S3.

Remarkably, similar two-phase regions are observed in experimentally deformed samples, although they were not reported in the initial publication[15]. Fig. 3d presents a transmission electron microscopy (TEM) image of compressed β-InSe, where the atomic arrangement transforms from a layered to a tetragonal structure. Enlarged views in Fig. 3d in the right-hand side provide a more detailed depiction of the layered hexagonal and of the tetragonal structures. The layered structure exhibits an interlayer spacing of 0.843 nm, which was confirmed as β-InSe with [120] orientation through calibration with electron diffraction patterns, as shown in Fig. 3f. On the other hand, the tetragonal structure displays an orthorhombic lattice, with slight differences along two directions. An analysis of electron diffraction patterns, as shown in Fig. 3f, revealed that the spacing between ($1\bar{1}0$) planes is approximately 2.7% larger than between (004) planes. Based on the atomic structure of the tetragonal phase shown in Fig. 3c, we can confidently identify it as a tetragonal phase with the [110] orientation. Furthermore, we acquired aberration-corrected electron microscopy images of the phase transformation interface. As shown in Fig. 3e, the orientation relationship between the product and parent phases observed in this image perfectly matches that of the simulations in Fig. 3b.

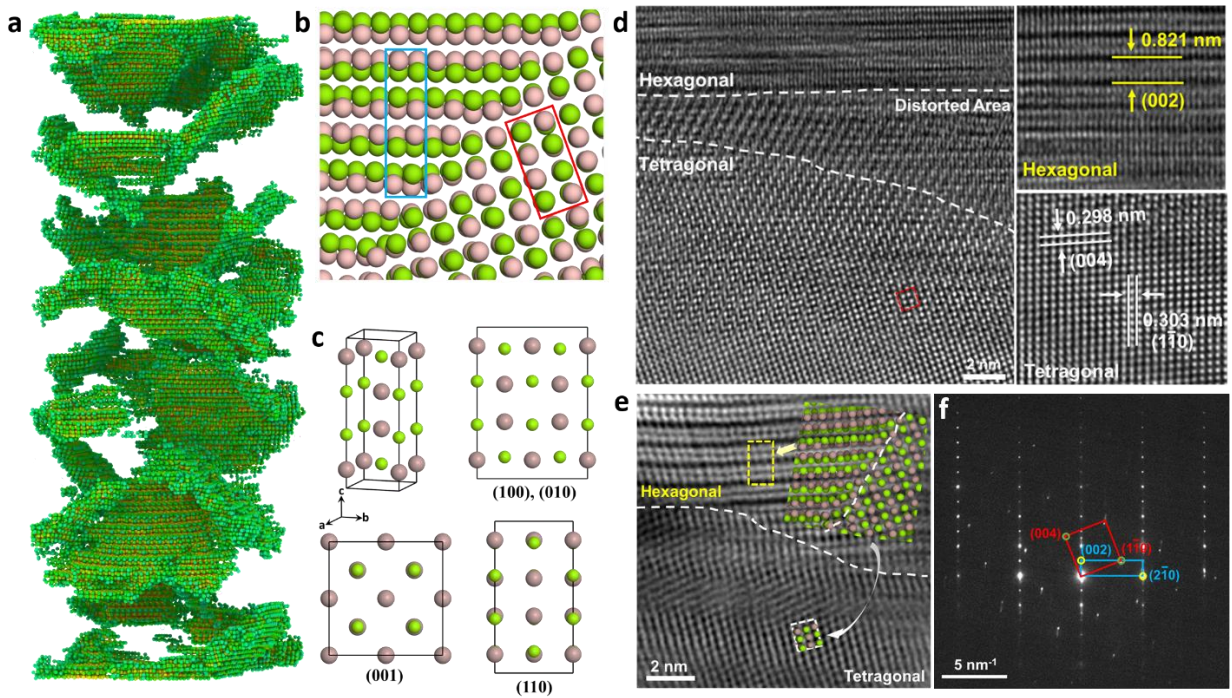

**Fig. 3| Phase transformation during InSe compression: numerical and experimental insights**. **a**, Nucleation of multiple tetragonal regions during the compression of hexagonal InSe (the parent phase was identified and removed thanks to its distinct coordination number). **b**, Mmagnified view of a selected two-phase interface. **c**, Single-cell structure of the tetragonal phase along with the atomic arrangement in the (100), (010), (001), and (110) planes. **d**, Regions of the two-phase interface discovered in an experimentally deformed sample, along with the enlarged views of layered hexagonal and tetragonal structures. **e**, Orientation relationship between the product and parent phases in the experiment and compares them with the

simulations. **f**, Electron diffraction pattern of **d**, featuring two patterns that correspond to the hexagonal layered phase with [120] direction and the tetragonal phase with [110] direction.

## Martensitic crystallographic relationships

As seen in Fig. 3b, the hexagonal and tetragonal phases adopt specific orientation relationships and create a coherent interface, both features being typical of martensitic transformations. Two relationships were observed in the simulations with different coherency axes: orientation relationship OR-1 with $[120]_H//[110]_T$ and orientation relationship OR-2 with $[100]_H//[100]_T$. Note that OR-1 corresponds to the experimental observation in Fig. 3d. We applied the phenomenological theory of martensitic transformations[43–45] to better understand the crystallography of these ORs.

By inputting the shape and dimensions of parent and product unit cells corresponding to both ORs (see Supplementary Materials S4 for details), we first confirmed that in both cases, the coherency axes noted above correspond to the smallest strain, thus likely accommodated elastically. We then predicted the habit planes that are shown as red lines in Figs. 4a and d on top of atomic structures extracted from the simulations. A perfect match is obtained, further confirming the martensitic nature of the transformation.

Figs. 4b and e illustrate the relative orientation between the parent hexagonal phase unit cell (H-cell) and the product tetragonal phase unit cell (T-cell) for both ORs. With OR-1, there is an approximate rotation of 20° of the product phase around the $[120]_H//[110]_T$ coherency axis, while with OR-2 the rotation is of 12° around the $[100]_H//[100]_T$. In order to further compare both ORs, we used the solid state nudged elastic band (ssNEB) method to compute the minimum energy path (MEP) of transformation between parent and product phases. By constructing different initial paths, we obtained the MEP for both ORs. The atomic structures corresponding to the states marked by red dots on the MEP curve are displayed as insets (Figs. 4c and f), revealing the transformation mechanism from the hexagonal to the tetragonal lattice. The atomic structure with maximum energy along the MEP is the transition state. For OR-1, we observed that atomic layers gradually transform from a "zigzag" arrangement to a linear one. In the transition state, new covalent bonds predominantly form between In and Se atoms on either side of the van der Waals layer due to compression. In the final state, the formation of covalent bonds between In and Se atoms within the van der Waals layer that are not adjacent leads to the formation of a tetragonal lattice. We determined the formation of these new bonds by calculating the differential charge density for all converged structures; for further details, please refer to the Supplementary Materials S5. The energy barrier is 288.4 meV per unit cell with OR-1.

OR-2 requires the relative sliding of atomic layers on the left and right sides of the hexagonal lattice along the "zigzag" direction, as indicated by the green and blue regions in the insets of Fig. 4f. The process of new bond formation is the same as with OR-1. A new bond is formed between In and Se atoms on either side of the van der Waals layer in the transition state and another new bond is formed between In and Se atoms within the van der Waals layer that are not adjacent. The energy barrier for OR-2 is determined to be 295.8 meV per unit cell, only slightly higher than for OR-1. This explains why in the simulations both lattice orientations are observed.

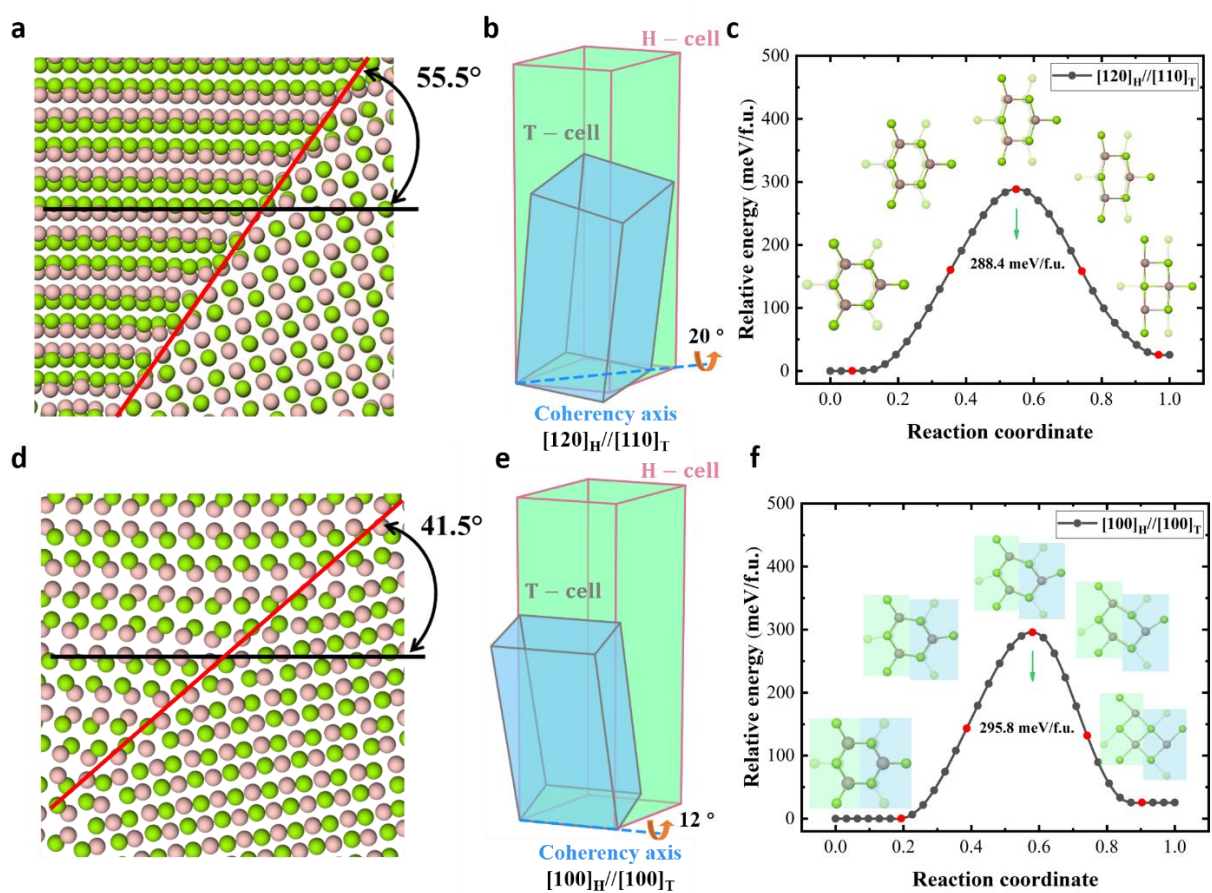

**Fig. 4|Crystallographic relationships during martensitic transformation from hexagonal to tetragonal phases. a,d**, Comparison of the predicted habit planes (red lines) from the martensitic theory with the actual phase transformation interface, revealing a precise alignment. **b,e**, Schematic diagrams illustrating the coherency axis and specific orientation relationship for each ORs. **c,f**, MEP curves of the ORs, with insets corresponding to the atomic structures of the states marked by red dots on the MEP curve.

## Impact of the tetragonal phase on the plasticity of InSe

To illustrate the impact of phase transformation on the plasticity of InSe, we carried out two comparative stress unloading processes on deformed samples, as depicted in Fig. 5a. We initially conducted stress unloading at the end of elastic deformation, observing a perfect overlap between the stress-strain curve during unloading (blue curve) and the elastic segment of the loading stress-strain curve (red curve). This observation confirms the absence of plastic deformation contribution in this stage. Subsequently, stress unloading was applied at the deformation termination, and the stress-strain curve (purple curve) exhibited a partial recovery of deformation. Nonetheless, a 12% residual deformation persisted when the stress was unloaded to zero, indicating a transition from elastic to plastic deformation during the stress drop stage of the loading stress-strain curve. Various states were selected during deformation marked by stars in Fig. 5a, providing a clear illustration of the nucleation and growth process of the new tetragonal phase, as demonstrated in Fig. 5b.

For inorganic semiconductors, their deformation manifests in two distinct stages. The initial stage is elastic, where deformation primarily occurs at the bond level, undergoing compressive and tensile deformation of recoverable bonds. The extent of deformation is contingent upon the stiffness of the bonds; for brittle materials, this deformation is typically minimal. Conversely, in the case of plastic van der Waals (vdW) materials, the weak interlayer interactions enable a substantial compressible space, resulting in a more pronounced amount of elastic deformation. As the elastic stage concludes, sustained strain application induces a structural change in the material upon reaching the critical stress point, entering the second stage. Brittle materials, characterized by high bond rigidity, undergo structural damage while simultaneously releasing elastic strain. In contrast, plastic vdW materials predominantly convert most of the elastic strain into irreversible plastic deformation through nucleation and growth of new phase. The relevant schematic diagram is shown in Fig. 5c.

Moreover, phase transformation introduces two additional positive effects on enhancing plastic deformation. Firstly, the formation of the new phase bridges several van der Waals layers, augmenting interlayer interaction strength as depicted in the top two figures in Fig. 5d. Additionally, the new phase effectively hinders the propagation of cracks along van der Waals layers, as illustrated in the bottom two figures in Fig. 5d. The synergistic effect of these two factors suppresses material cleavage, resulting in unprecedented ductility.

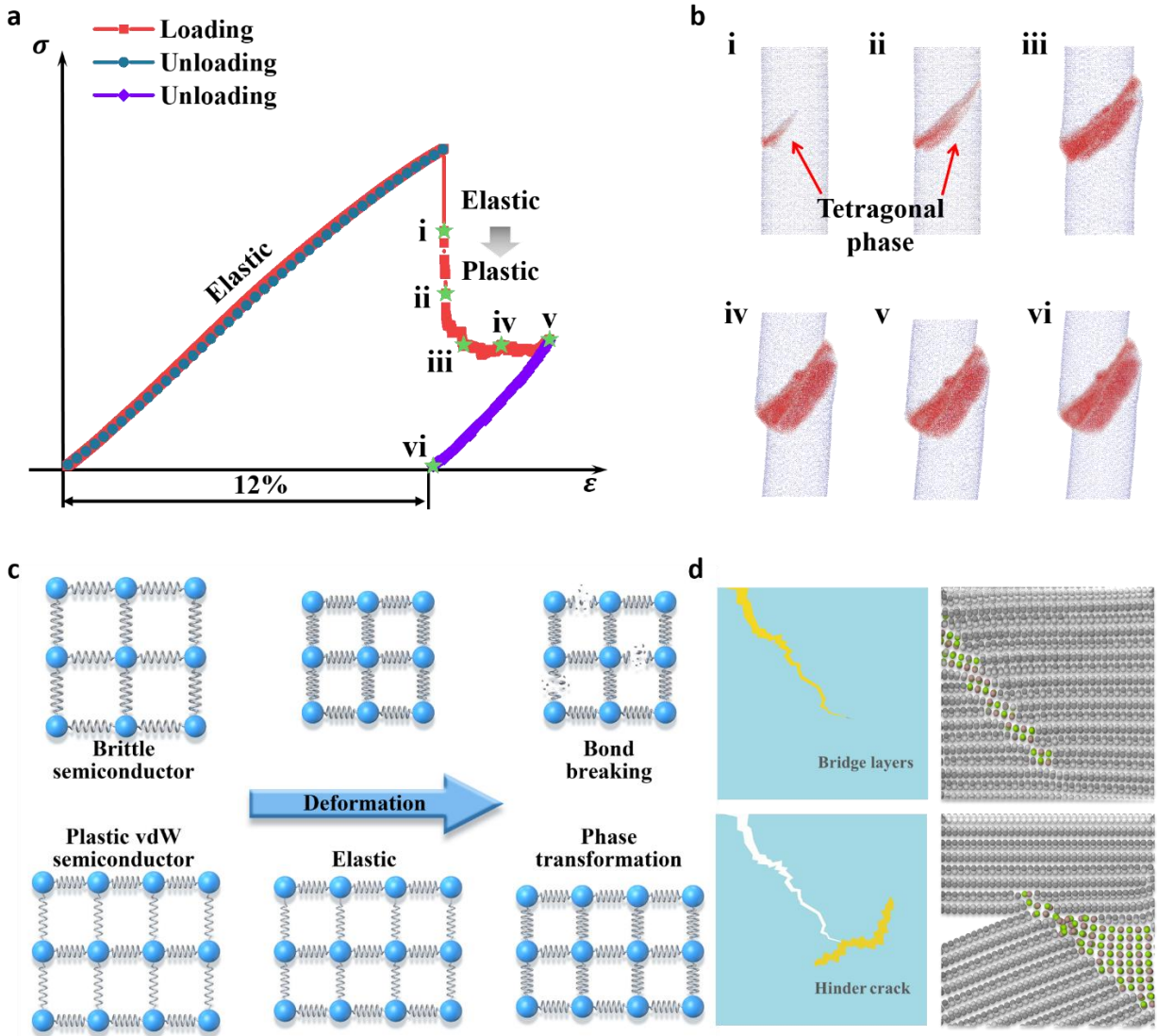

**Fig. 5| Phase transformation enhances the plasticity of InSe. a**, Strain-stress curves during loading and unloading. **b**, Illustration of the nucleation and growth process of the new tetragonal phase at the selected states in **a**. **c**, Schematic diagram of bond state change during deformation for both brittle semiconductor and plastic vdW semiconductor. **d**, Visualization of the bridging effect on van der Waals (vdW) layers and inhibition of crack propagation along the vdW layers induced by the new phase.

## Conclusions

In summary, this study, facilitated by a machine-learned deep potential, has unveiled a previously-unknown martensitic phase transformation in 2D vdW crystal InSe upon compression, which is key to the exceptional plasticity. The crystallographic characteristics of this transformation have been elucidated through a synergy of simulation, theoretical analysis, and experimental validation. Confirming the existence of a stress-induced phase transformation from a low-density hexagonal structure to a high-density tetragonal phase in the deformed samples. This transformation does not require breaking strong covalent bonds while forming bonds between layers which fully harnessed the out-of-plane degrees of freedom in these materials. This discovery provides new insight into the mechanical behavior and mechanism of 2D vdW materials, which will advance the development of more versatile and deformable inorganic materials for a wide range of applications.

# Methods

### DFT calculations

Density functional theory (DFT) calculations were performed using the pseudopotential plane-wave method implemented in the Vienna *ab initio* Simulation Package (VASP) code[46]. To account for van der Waals interactions, we applied the Perdew-Burke-Ernzerhof (PBE) generalized-gradient approximation (GGA) functional[47,48] with the optB86b-vdW functional[49]. The projector augmented-wave (PAW) method[50] was used to describe interactions between core and valence electrons, with the following valence configurations: In-$4s^2 3d^{10} 4p^1$, Se-$4s^2 4p^4$. The electronic wave functions were expanded in a plane-wave basis set with a kinetic energy cutoff (ENCUT) of 650 eV, and a Monkhorst-Pack k-point sampling with grid spacing (KSPACING) of 0.13 Å$^{-1}$. Convergence was reached on total energy, force, and virial tensor below 1 meV/atom, 10 meV/Å and 1 meV/atom, respectively.

### DP constructions

The Deep Potential (DP) model underwent training through the construction of a DFT dataset, which was meticulously constructed via a concurrent learning approach known as the Deep-Potential Generator (DP-GEN). The process starts with an initial dataset, and DP-GEN proceeds through a series of iterative steps encompassing training, exploration, and labeling. In the training process, four models are trained using the same training dataset and hyper-parameters but different random seeds. In the exploration process, the configuration space of InSe is sampled through MD simulations using the different DP models. These simulations were conducted using the LAMMPS package[51], which is linked to the DeePMD-kit. The configurations are evaluated using the four DP models initially trained. If the maximum deviation of atomic forces within a configuration falls within predefined upper and lower bounds, we classify the configuration as a candidate. These candidate configurations are then subjected to label assignment, which involves calculating essential attributes such as energy, force, and virial tensor using DFT. Subsequently, these labeled candidates are incorporated into the training dataset. For a more comprehensive understanding of the construction of the DP dataset and the model training process, detailed technical information is provided in Supplementary Materials S2. This potential is freely available on the AIS Square website[52].

**MD simulation**

The MD simulations were carried out using the LAMMPS[51] package, employing a timestep of 1 femtosecond. The systems were firstly relaxed in the isothermal-isobaric (NPT) and canonical ensembles (NVT) and then deformed by changing the length of the simulation box with energy minimization employed to achieve equilibrium. A force threshold of 1.0E-2 eV/ Å was selected to determine convergence of the calculation.

**ssNEB**

A comprehensive description of the ssNEB principles and implementation can be found in Ref. 53. The ssNEB calculations were performed with the widely used Atomic Simulation Environment[54] (ASE) Python library and the Transition State library for ASE (TSASE). For both ORs, we used 16 atom-cells of the respective starting and final points, with particular attention to the correspondence of atomic indexes between initial and final configurations. In both cases, initial and final image positions and cells were relaxed using the Broyden-Fletcher-Goldfarb-Shanno[55] (BFGS) algorithm implemented in ASE to ensure a maximum atomic force of 1.0E-10 eV/ Å. We then performed climbing image ssNEB with 30 intermediate images, a spring constants of 5.0 eV/ Å$^2$ and the Fast Inertial Relaxation Engine[56] (FIRE) optimization algorithm, with a stopping criterion reached when no atomic forces on any image was greater than 1 meV/ Å.

**Crystal growth and material preparation**

InSe polycrystal was firstly prepared before crystal growth. Raw materials of In and Se with a nonstoichiometric molar ratio of 52: 48 were sealed in a quartz tube under an argon atmosphere at a pressure of ~1.0E-3 Pa. Then, the tube was placed in a furnace and slowly heated to 1273 K. After holding at this temperature for 10 h and cooling to room temperature, InSe polycrystal was obtained. For crystal growth, a modified Bridgman method was adopted. The polycrystal was re-sealed and put in a two-zone vertical furnace. The upper zone was set to 973 K for melting, and the lower zone was set as 573 K. The crystal growth was carried out with a rate of 0.5 mm/h for about one week.

**Structure characterization**

The atomic structure characterization of InSe crystals before and after compression was conducted by transmission electron microscopy (TEM, TALOS F200X, Thermofisher; JEM-ARM200F, JEOL). The observed TEM samples were cut perpendicular to the compression surface before and after compression using focused-ion-beam (FIB, GAIA3, TESCAN).

**Data availability**

The authors declare that the main data supporting the findings of this study are contained within the paper and its associated Supplementary Materials. Source data are provided in this paper.

**Code availability**

LAMMPS, DP-GEN, DeepMD-kit, PHONOPY, and ShengBTE are free and open-source codes available at https://lammps.sandia.gov, https://deepmodeling.com, https://phonopy.github.io/phonopy, and https://www.shengbte.org, respectively. ssNEB code is available from the authors upon request.

## Acknowledgments

This work is supported by the National Natural Science Foundation of China (T2122013, 51790494, U2330401). The authors thank Dr. Dylan Bissuel for the ssNEB calculations. B. X. is grateful to Prof. Dr. Yun-Jiang WANG for valuable discussions. Y.D. thanks Dr. Xiaoyang Wang for helpful discussions on the construction of the machine-learned potential.

## Author contributions

D.R., T.W., and B.X. contributed to the conception, design, and management of the work. Y.S. contributed to the training of the potential and MD simulations. Y.M. contributed to the preparation of the samples and compress test. All authors participated in the interpretation of data, the writing, and the preparation of the manuscript. Y.S. and Y.M. contributed equally to this work.

## Competing interests

The authors declare no competing interests.

## Additional information

Supplementary information. The online version contains supplementary material available at [URL will be added by the publisher]

# Supplementary materials for

# Van der Waals semiconductor InSe plastifies by phase transformation

Yandong Sun[1,±], Yupeng Ma[2,±], Jin-yu Zhang[3,5], Tian-Ran Wei[2*], Xun Shi[2,4],

David Rodney[3*], Ben Xu[1,5*]

[1]Graduate School of China Academy of Engineering Physics, Beijing 100193, China

[2]State Key Laboratory of Metal Matrix Composites, School of Materials Science and Engineering, Shanghai Jiao Tong University, Shanghai 200240, China

[3]Univ. Lyon, CNRS, Université Claude Bernard Lyon 1, Institut Lumière Matière, UMR5306, Villeurbanne 69622, France

[4]State Key Laboratory of High Performance Ceramics and Superfine Microstructure, Shanghai Institute of Ceramics, Chinese Academy of Sciences, Shanghai 200050, China

[5]AI for Science Institute, Beijing 100080, China

[±]These authors contributed equally to the work.

E-mail: tianran_wei@sjtu.edu.cn; david.rodney@univ-lyon1.fr; bxu@gscaep.ac.cn

## S1 DENSITY FUNCTIONAL THEORY CALCULATIONS

The convergence tests on total energy for β-InSe and γ-InSe are illustrated in Fig. S1. In different types of computations, the criteria for energy and force convergence may differ due to the varying levels of convergence challenges. To offer clarity, the specific convergence criteria for each will be detailed in the relevant sections.

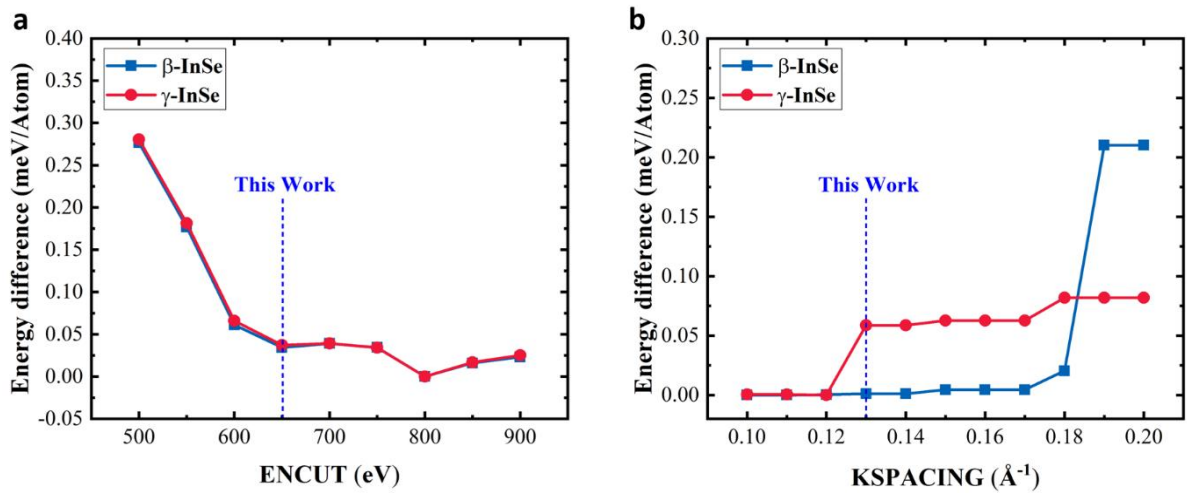

**Fig. S1|Convergence tests on total energy in VASP[1] calculations for the β- and γ-InSe**. Total energies as a function of **a** kinetic energy cutoff (ENCUT) and **b** grid spacing of k-point sampling (KSPACING), relative to the results obtained using more stringent parameters (ENCUT = 900 eV and KSPACING = 0.10 Å$^{-1}$)

## S2 GENERATION OF THE DEEP POTENTIAL MODEL

### A. Protocol for the construction of the bulk dataset

*Initial Dataset*: To kickstart the DP-GEN[2] process and ensure stable sampling during its iterative evolution, an initial step involved the training of four DP models. These models were established based on a preliminary dataset that contains two types of configurations:

1. *Locally Optimized Configurations*: β- and γ-InSe prototypes.

2. *Locally Perturbed Configurations*: Following the optimization of local configurations, further perturbations were introduced with the mean perturbation distance of atoms set to 0.03 Å, and the mean perturbation fraction of cell vectors set to 3.0%. For each structure at each pressure point, a total of 4 perturbations were carried out and subsequently integrated into the preliminary dataset.

In total, the preliminary dataset included 200 configurations. The labels associated with these configurations, including energies, forces, and virial tensors, were derived through high-precision single-point DFT calculations with energy and force convergences less than $10^{-9}$ eV and $10^{-5}$ eV/Å, respectively.

*Initialization of Models*: To ensure model diversity, various random seeds were employed to initialize the model parameters. In order to enhance the representational capacity of the DP model, we incorporated a three-body embedding descriptor, seamlessly integrating it with the original smooth DP descriptor. You can find the technical details of these descriptors in Ref. 3. The embedding network of size (60, 120, 240) follows a ResNet-like architecture[4]. The fitting network is composed of three layers, each containing 240 nodes.

*Training Process*: During training, a loss function was defined as follows (Eq. 1), which quantifies the model's accuracy based on disparities between DFT-calculated energy ($\hat{E}^k$), forces ($\hat{F}_{i\alpha}^k$), and the virial tensors ($\hat{\Xi}_{\alpha\beta}^k$), as compared to their predictions by the DP model.

$$\mathcal{L} = \frac{1}{|\mathcal{B}|}\Sigma_{k\in\mathcal{B}} \left( p_\epsilon \frac{1}{N}|\hat{E}^k - E^k|^2 + p_f \frac{1}{3N}\Sigma_{i\alpha}|\hat{F}_{i\alpha}^k - F_{i\alpha}^k|^2 \right.$$
$$\left. + p_\xi \frac{1}{9N}\Sigma_{i\alpha}|\hat{\Xi}_{\alpha\beta}^k - \Xi_{\alpha\beta}^k|^2 \right), \quad (1)$$

In this equation, $\mathcal{B}$ represents a subset of the data, and $|\mathcal{B}|$ is the batch size. Each training data includes a configuration, comprising atom coordinates, box basis vectors, and element types, alongside their corresponding labels. Prefactors ($p_\epsilon$, $p_f$, $p_\xi$) are a set of hyper-parameters that dictate the relative importance of energy, force, and the virial tensor throughout training. These prefactors adapt gradually in line with the learning rate ($r_l(t)$), which exponentially decays with the training step ($t$) as

$$r_l(t) = r_l^0 k_d^{\frac{t}{t_d}}, \quad (2)$$

where $r_l^0$ is the learning rate at the beginning of the training, $t_d$ denotes the typical timescale of the learning rate decay and $k_d$ denotes the decay rate. The prefactors evolve with the learning rate as

$$p_\alpha(t) = p_\alpha^{\text{limit}} \left[1 - \frac{r_l(t)}{r_l^0}\right] + p_\alpha^{\text{start}} \left[\frac{r_l(t)}{r_l^0}\right], \alpha = \epsilon, f, \text{or } \xi, \quad (3)$$

where $p_\alpha(t)$ is either of the three prefactors ($p_\epsilon$, $p_f$, $p_\xi$) at training step ($t$). $p_\alpha^{\text{start}}$ and $p_\alpha^{\text{limit}}$ are the prefactors at the beginning and at an infinitely small learning rate, respectively. During DP-GEN iterations, we set the following values: $p_\epsilon^{\text{start}} = 0.02$, $p_\epsilon^{\text{limit}} = 2.00$, $p_f^{\text{start}} = 1000.00$, $p_f^{\text{limit}} = 1.00$, $p_\xi^{\text{start}} = 0.00$ $p_\xi^{\text{limit}} = 0.00$, $t_d = 2000$, $k_d = 0.95$, and $r_l^0 = 1 \times 10^{-3}$. The total number of training steps conducted amounted to 400,000.

*Exploration Process*: In each iteration, the exploration involved navigating the configuration space using molecular dynamics (MD). We employed the LAMMPS package[5] with DeePMD-kit[3] support for isothermal-isobaric (NPT) Deep Potential molecular dynamics

(DPMD) simulations, lasting up to 60 picoseconds, with a timestep of 1 femtoseconds. Table S1 provides details regarding the pressure and temperature setting for DPMD simulations.

During DPMD simulations, we assessed the force prediction error by comparing forces among the four independently trained DP models. If the maximum deviation of atomic forces in a frame within the MD trajectory fell within specified upper and lower bounds, that frame was considered a candidate configuration. Up to 100 candidate configurations were labeled through DFT calculations in each DP-GEN iteration. The upper and lower bounds were adjusted relative to the pressure and are documented in Table S1. As the number of iterations increased, the deviation of force predictions among independently trained DP models progressively diminished. DP-GEN convergence was reached when 99.9% of explored configurations exhibited deviations in force predictions below the lower bound.

***Labeling Process***: The energy, force, and virial tensor labels for the candidate configurations were determined through DFT as described in Table S1. Subsequently, these candidate configurations and their associated labels were seamlessly integrated into the dataset for the next model training iteration.

**B. Protocol for the construction of the surface dataset**

After achieving convergence in the aforementioned model iteration process, we successfully obtained a Deep Potential (DP) model that can simulate the relevant properties of bulk materials. However, recognizing the pivotal role of surfaces in defect formation during deformation processes, we introduced training data from surface configurations to fine-tune the model. The methodology for acquiring the surface dataset parallels that of bulk materials. Initially, we obtained locally optimized surface configurations, including the (120) and (100) facets of the β-phase, supplemented with five different thicknesses of vacuum layers. Subsequently, we employed DP-GEN to undergo a series of iterative processes until reaching convergence. In each DP-GEN iteration, we labeled up to 90 candidate configurations through DFT calculations. Due to the relative convergence challenge for DPGEN in handling surface data, we lowered the energy convergence criteria to $10^{-5}$ eV for single-point DFT calculations. The upper and lower bounds were adjusted relative to the pressure and are documented in Table S1.

**Table S1.**

The number of atoms, pressure, temperature, and lower and upper bound of deviations of force predictions in the DP-GEN iterations.

| Structures | Number of Atoms | Sampling Pressure (GPa) | Temperature (K) | Lower Bound (eV/Å) | Upper Bound (eV/Å) |
|---|---|---|---|---|---|
| β-InSe-bulk | 72 | 0-50 | 0-1300 | 0.08 | 0.20 |
| γ-InSe-bulk | 72 | 0-50 | 0-1300 | 0.08 | 0.20 |
| β-InSe-(120)-surface | 8 | 0 | 0-300 | 0.05 | 0.12 |
| β-InSe-(100)-surface | 8 | 0 | 0-300 | 0.05 | 0.12 |

## C. Protocol for the construction of the specific dataset

*Deformation dataset*: We subjected the supercell of the β-phase to continuous compressing or stretching deformations along three directions: [120], [100], and [001]. In this process, one direction was held fixed while the other two were allowed to vary, and DFT structural optimizations were conducted. Moreover, we introduced shear strains by modifying the angle between the basic vectors [120] and [100] within the β-phase, following which the angle was fixed and DFT structural optimization was performed. Throughout these structural optimization processes, the convergence criteria for energy and force were set at $10^{-7}$ eV and $10^{-2}$ eV/Å, respectively. These configurations were then incorporated into the training dataset, with detailed information provided in Table S2.

*Cleavage energy dataset*: We modified the spacing between the van der Waals layers within the β-phase to simulate the cleavage process. Single-point DFT calculations were performed for these adjusted configurations, and the resulting data was subsequently integrated into the training dataset, see Table S2 for details.

*GSFE dataset*：We conducted calculations to determine the generalized stacking fault energy (GSFE) for six potential slip paths within the β-phase. These paths encompass both intralayer and interlayer slip, involving the typical low-index planes for the hexagonal crystal system—namely (120), (100), and (001)—as well as the typical slip directions [120], [100], and [001]. Furthermore, GSFE surface calculations were carried out specifically within the van der Waals layers of the β-phase. The resulting data from these analyses was subsequently integrated into the training dataset, see Table S2 for details.

**Table S2.**

The construction of the specific dataset for fine-tuning the DP model

| Specific dataset | Structures | Number of Atoms | Number of data |
|---|---|---|---|
| *Deformation* | Compress-[120] | 64 | 24 |
| | Tensile-[120] | 64 | 19 |
| | Compress-[100] | 64 | 25 |
| | Tensile-[100] | 64 | 15 |
| | Compress-[001] | 64 | 20 |
| | Tensile-[001] | 64 | 19 |
| | Shear-[120][100] | 72 | 13 |
| *Separation energy* | vdW layer | 72 | 40 |
| *GSFE paths* | (001)[100] | 16 | 11 |
| | (001)[120] | 16 | 21 |
| | (100)[001] | 32 | 21 |
| | (100)[120] | 32 | 11 |
| | (120)[001] | 32 | 21 |
| | (120)[100] | 32 | 11 |
| *GSFE surface* | vdW layer | 72 | 121 |

**D. Validation**

After preparing the training data, a prolonged training was conducted based on this dataset with training parameters: $p_\epsilon^{\text{start}} = 0.02$, $p_\epsilon^{\text{limit}} = 2.00$, $p_f^{\text{start}} = 1000.00$, $p_f^{\text{limit}} = 2.00$, $p_\xi^{\text{start}} = 0.005$ $p_\xi^{\text{limit}} = 0.1$, $t_d = 60000$, and $r_l^0 = 1 \times 10^{-3}$. The total number of training steps was $1.2 \times 10^7$.

The DP model reliability underwent three validation processes: testing against the training dataset, examination of lattice parameters, and assessment of thermodynamic properties, encompassing phonon dispersion, thermal conductivity, mode Grüneisen parameter, and scattering rate.

*Testing Against the Training Dataset*: We conducted a comparative analysis between the energy and atomic force computed using the DP model and those derived from DFT for a total of 19,423 configurations within the final training dataset. Our analysis demonstrated a strong agreement between the DP and DFT results, with a mean absolute error of 7.3 meV per atom for energy and 0.07 eV/Å for atomic force. These results are presented in Fig. S2.

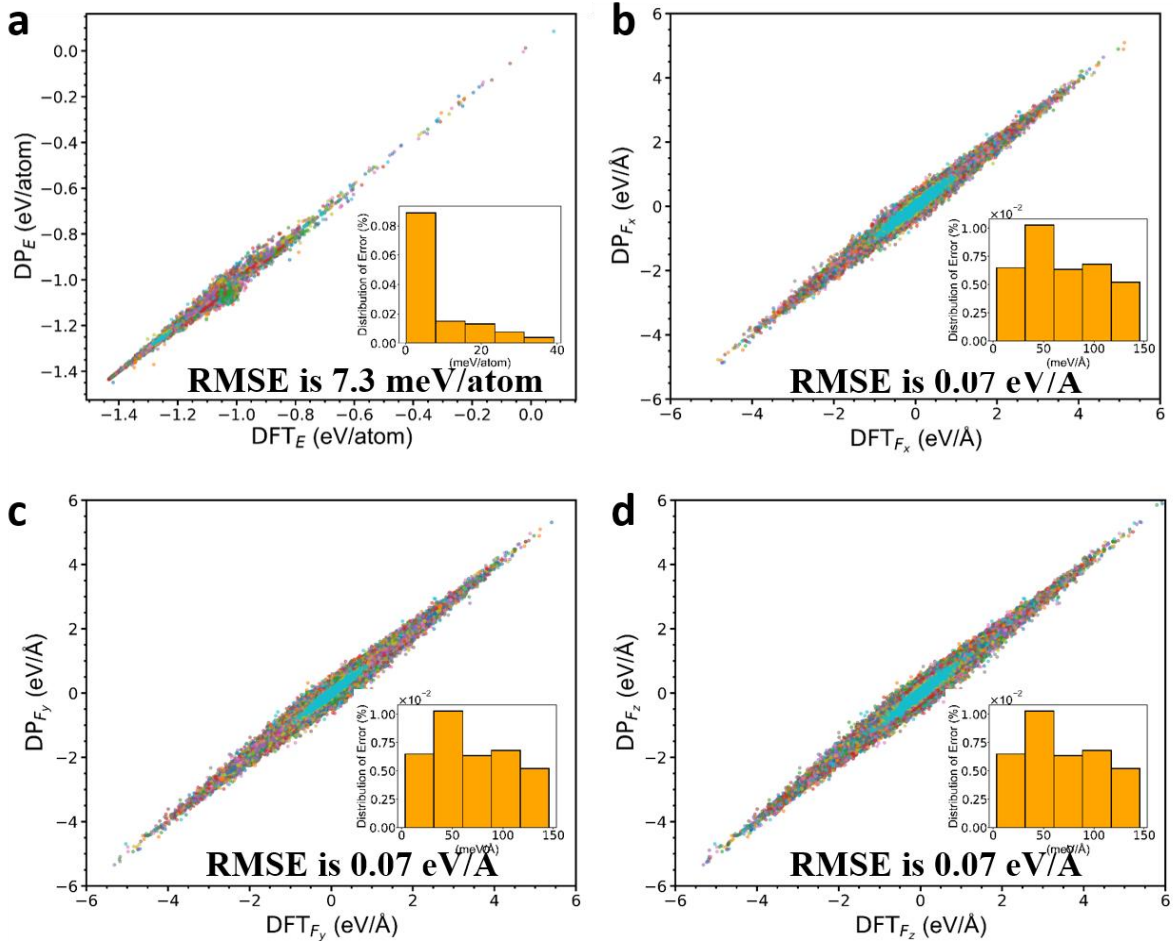

**Fig. S2| Comparison of energy and force between DP and DFT results. a**, Energy results. **b-d**, Force result with components in the *x*, *y*, and *z* directions, respectively.

*Examination of Lattice Parameters*: Both DP model and DFT were utilized independently to calculate the lattice constants for both β- and γ-InSe. The outcomes exhibited a notably high level of consistency between the two methods, as evidenced in Table S3, emphasizing the accuracy and reliability of the DP model.

**Table S3.**

The lattice parameters for the β- and γ-InSe using DP model and DFT method.

|  |  | $a = b$ (Å) | $c$ (Å) | $\alpha = \beta$ (°) | $\gamma$ (°) |
|---|---|---|---|---|---|
| β-InSe | DFT | 4.053 | 16.914 | 90.0 | 120.0 |
|  | DP | 4.063 | 16.764 | 90.0 | 120.0 |
| γ-InSe | DFT | 4.058 | 25.132 | 90.0 | 120.0 |
|  | DP | 4.067 | 24.928 | 90.0 | 120.0 |

*Thermal Dynamic Properties*: Due to InSe's potential as an energy conversion material, its thermal properties are of paramount importance for industrial applications. Consequently, we employed the DP model to compute the phonon dispersion, thermal conductivity, mode

Grüneisen parameter, and phonon scattering rate of β-InSe. These results were then compared to corresponding outcomes obtained using DFT, aiming to validate the accuracy and reliability of the DP model. For the calculation of the phonon dispersion, we utilized the finite displacement method as implemented in the PHONOPY code[6]. This method involves computing the phonon dispersion by determining the second-order force constants. Regarding thermal conductivity, mode Grüneisen parameter, and phonon scattering rate, we employed the Boltzmann transport equation to obtain these properties. This process required inputting the second-order and third-order force constants into the Shengbte software[7]. The comparison of results, as illustrated in Fig. S3, clearly demonstrates a high level of agreement between the thermal properties computed using the DP model and those obtained through DFT calculations. The systematic benchmark shows that the DP model has excellent DFT-level accuracy, and is capable of predicting thermal properties.

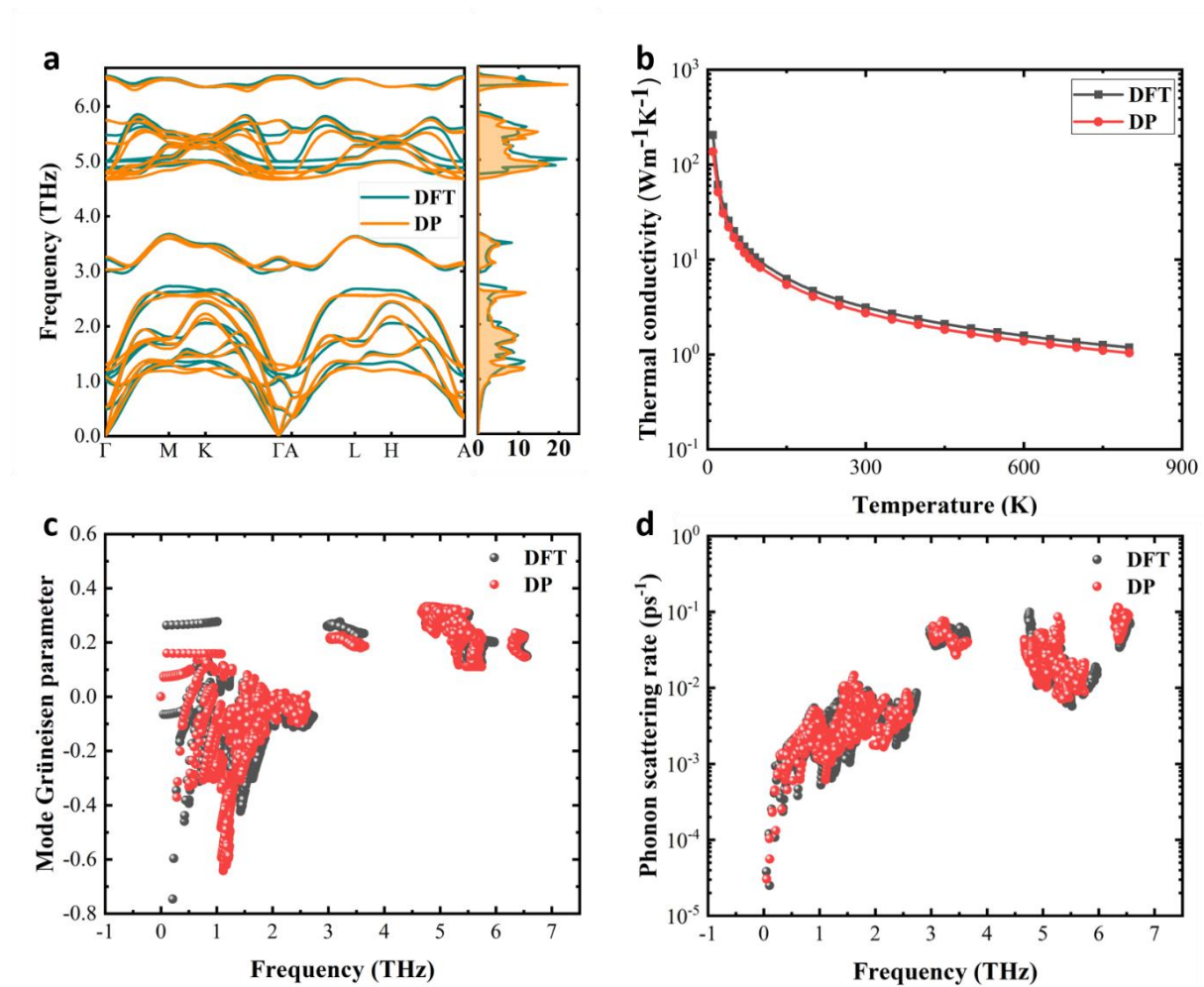

**Fig. S3| Comparative analysis of the thermal dynamic properties of β-InSe using DP and**

**DFT methods. a**, Phonon dispersion. **b**, Thermal conductivity. **c**, Mode Grüneisen parameter. **d**, Phonon scattering rate.

## S3 LATTICE PARAMETERS AND PHONON DISPERSION OF TETRAGONAL PHASE

We calculate the lattice parameters for the tetragonal in DFT. The lengths of basis vectors ***a*** and ***b*** are equal, measuring 4.183 Å, whereas the length of basis vector ***c*** is 11.517 Å. The angles among the basis vectors ***α***, ***β***, and ***γ*** are 90°.

We calculate the phonon dispersion of the tetragonal phase, see Fig. S4, imaginary frequency phonon in the zone center ($\Gamma$), and the zone-boundary (X) wavevector was observed, which is the same as the tetragonal phase for SnS and SnSe[8]. However, the imaginary frequency phonons can be eliminated by adding pressure up to 3 GPa to the system.

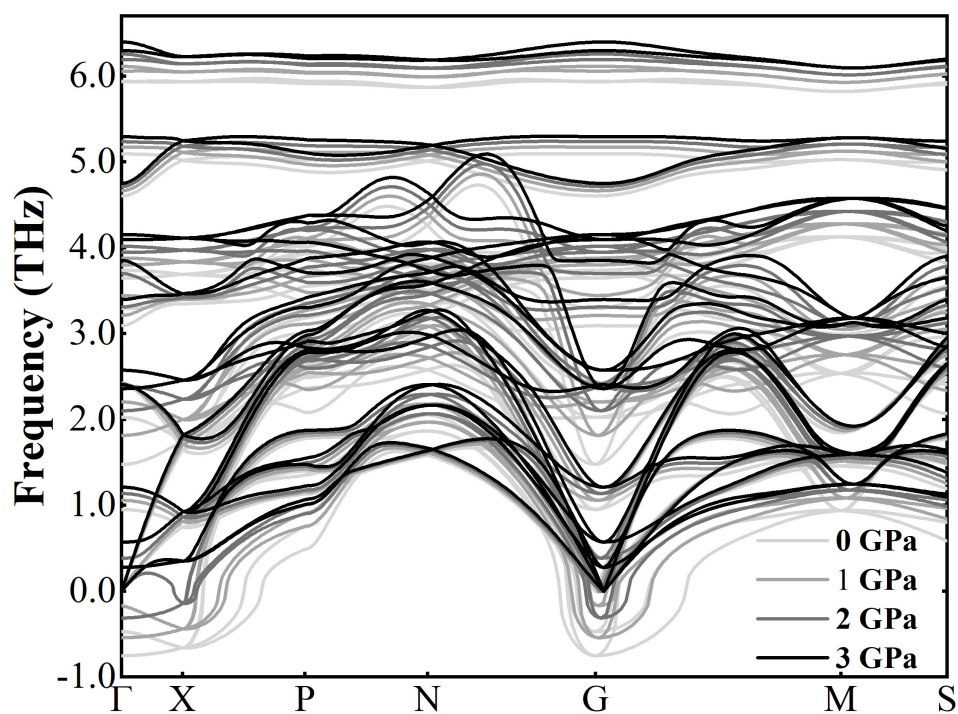

**Fig.S4| Phonon dispersion of tetragonal phase under different pressure.** The imaginary frequency phonons are eliminated by adding pressure up to 3 GPa.

## S4 THEORY OF MARTENSTITE TRANSFORMATION

The hexagonal-to-tetragonal transformation can be analyzed using the Phenomenological Theory of Martensitic Transformations (PTMT)[9–12]. The aim of this theory is to identify the habit plane and shape strain vector of the phase transformation by writing the latter as an invariant plane strain (IPS). Since the strain tensor [*F*], which geometrically characterizes the transformation by relating the unit cells of the parent and product phases, is itself generally not an IPS, it is usually assumed that the transformation is preceded by a lattice invariant shear, due to dislocation slip and/or twinning[9,10]. However, in the present case, neither the experiments, nor the simulations indicate the presence of dislocations nor twins, which might be related to the small scale of the experimental and numerical samples. We thus assume that the interface between transforming phases is fully coherent and elastically strained along a given direction contained in the interface, which will be the correspondence axis common to both the parent and product phases.

### OR-1

We use an orthorhombic cell of β-InSe to conduct the prediction, see Fig. S5 for the initial and final cells. Here already diagonal since both unit cells are orthorhombic.

$$[F] = \begin{bmatrix} \frac{5.92}{7.02} = 0.84 & 0 & 0 \\ 0 & \frac{5.92}{4.05} = 1.46 & 0 \\ 0 & 0 & \frac{11.50}{16.91} = 0.68 \end{bmatrix} \quad (4)$$

The coherency axis is thus $[120]_H = [110]_T$, with a coherency strain along this axis of -15.67%. The normal to both possible habit planes and associated strain vectors are:

$$n = [0.0 \quad -0.82 \quad 0.57] \quad s = [0.0 \quad -0.44 \quad -0.65]$$
$$n = [0.0 \quad 0.82 \quad 0.57] \quad s = [0.0 \quad 0.44 \quad -0.65]$$

They are related by a mirror symmetry with respect to the (001) plane, as observed in the simulations.

### OR-2

$$[F] = \begin{bmatrix} \frac{8.37}{7.02} = 1.19 & 0 & 0 \\ 0 & \frac{4.19}{4.05} = 1.03 & 0 \\ 0 & 0 & \frac{11.50}{16.91} = 0.68 \end{bmatrix} \quad (5)$$

The coherency axis is thus $[100]_H = [100]_T$ with a coherency strain along this axis is 3.36%. The normal to both possible habit planes and associated strain vectors are:

$$n = [-0.66 \quad 0.0 \quad 0.75] \quad s = [-0.23 \quad 0.0 \quad -0.46]$$

$$n = [0.66 \quad 0.0 \quad 0.75] \quad s = [0.23 \quad 0.0 \quad -0.46]$$

They are related by a mirror symmetry with respect to the (001) plane, as observed in the simulations.

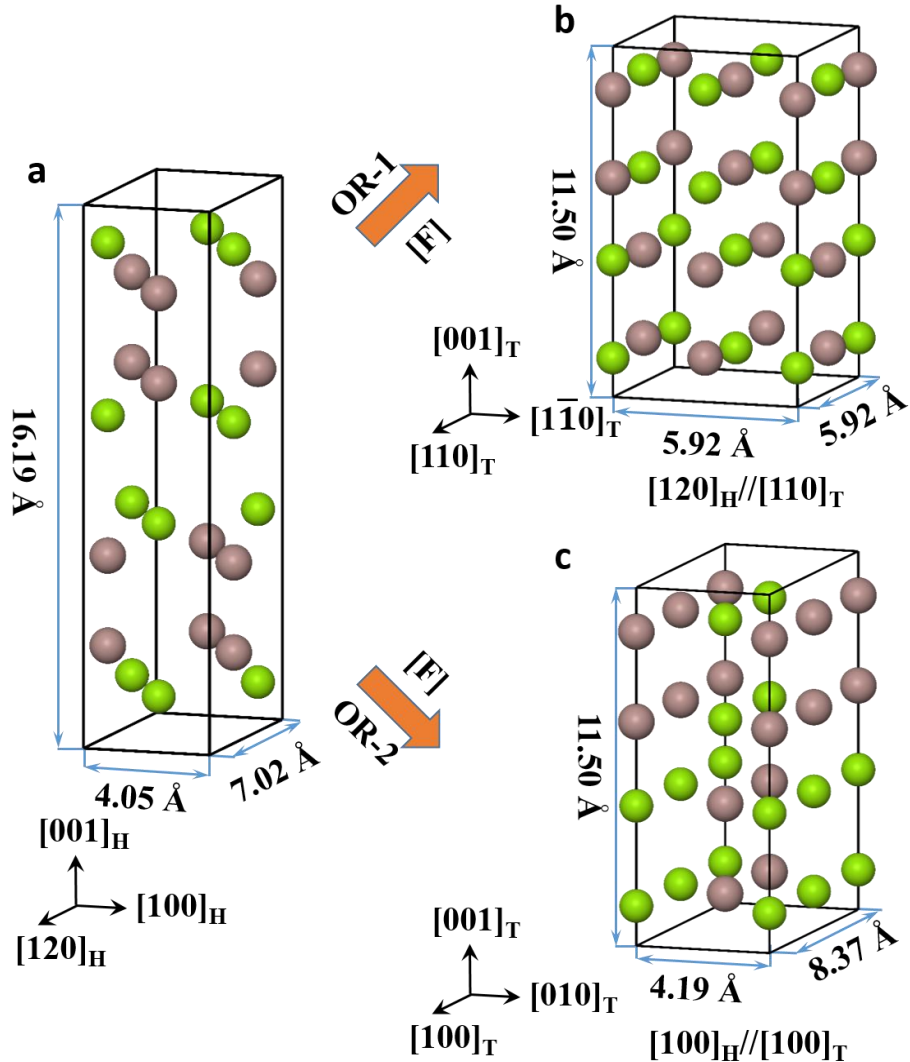

**Fig.S5| The initial and final unit cells for PTMT prediction**. **a**, Initial orthorhombic cell of β-InSe. **b,c**, Cells of tetragonal InSe with different orientation.

## S5 FORMATION OF NEW BONDS DURING PHASE TRANSORMATION

In our investigation of the orientation relationship (OR) between the tetragonal product phase and the parent hexagonal phase using solid state nudged elastic band (ssNEB) analysis, we obtained a series of atomic configurations along the minimum energy path (MEP). We calculated the differential charge density for the saddle configuration, its preceding and subsequent configurations, as well as the final three configurations along the MEP, as depicted in Figs. S6 and S7.

The result reveals that near the saddle configuration, new covalent bonds predominantly form between In and Se atoms across the van der Waals interlayers due to compression. This phenomenon is clearly observable through the movement of charge density, as indicated by the red arrows in panels **a**, **b** and **c** in Figs. S6 and S7. For the last three configurations along the MEP, we primarily observe the formation of covalent bonds between In and Se atoms that are not adjacent within the layers, corresponding to the new electron distribution positions illustrated by green arrows in panels **d**, **e,** and **f** in Figs. S6 and S7.

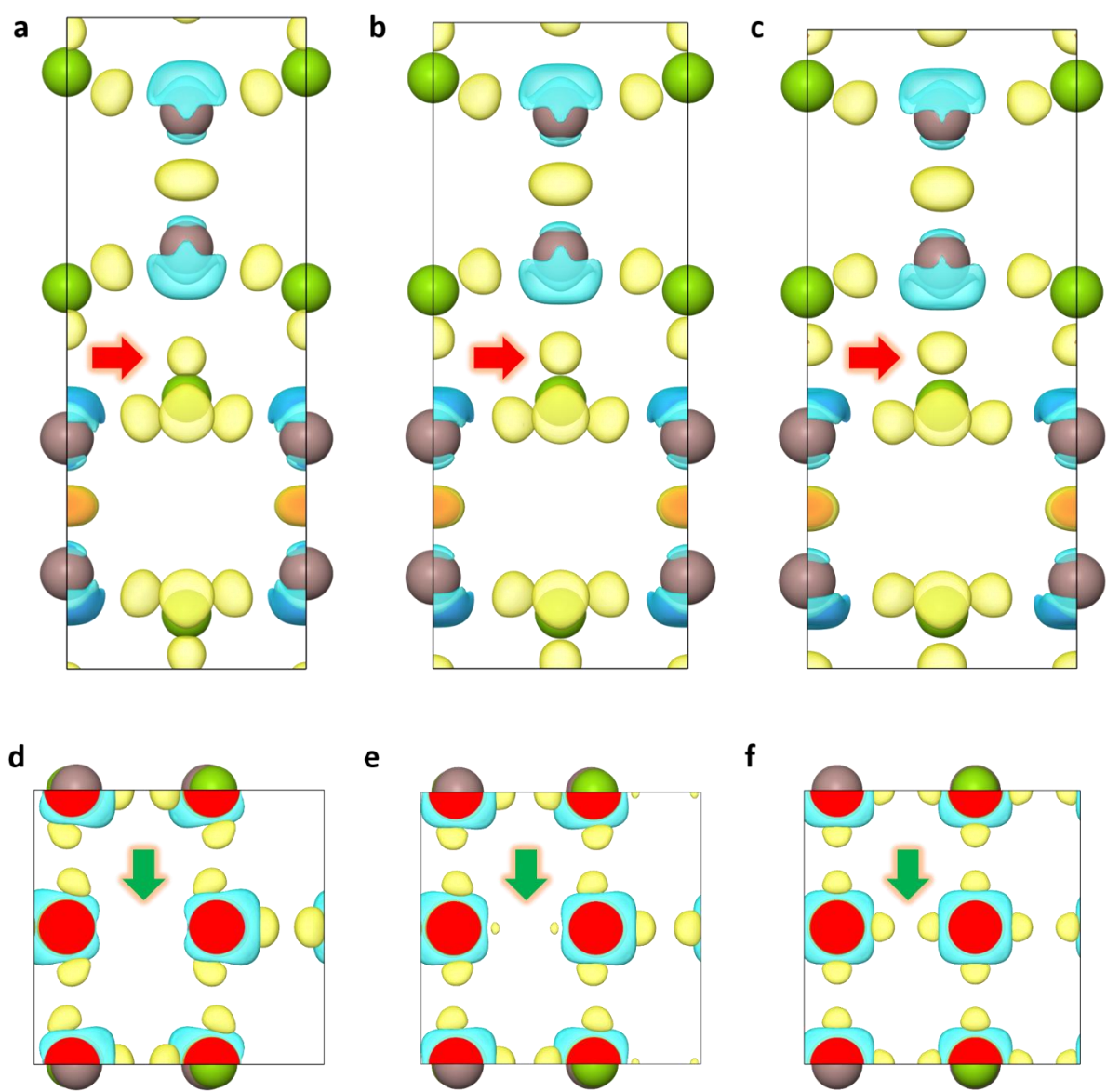

**Fig. S6| Differential charge density for selected configurations in the MEP for OR-1**. **a-c**, Differential charge density of the configurations before the saddle point, at the saddle point, and after the saddle point along the MEP. **d-f**, Differential charge density of the

configurations for the last three configurations along the MEP.

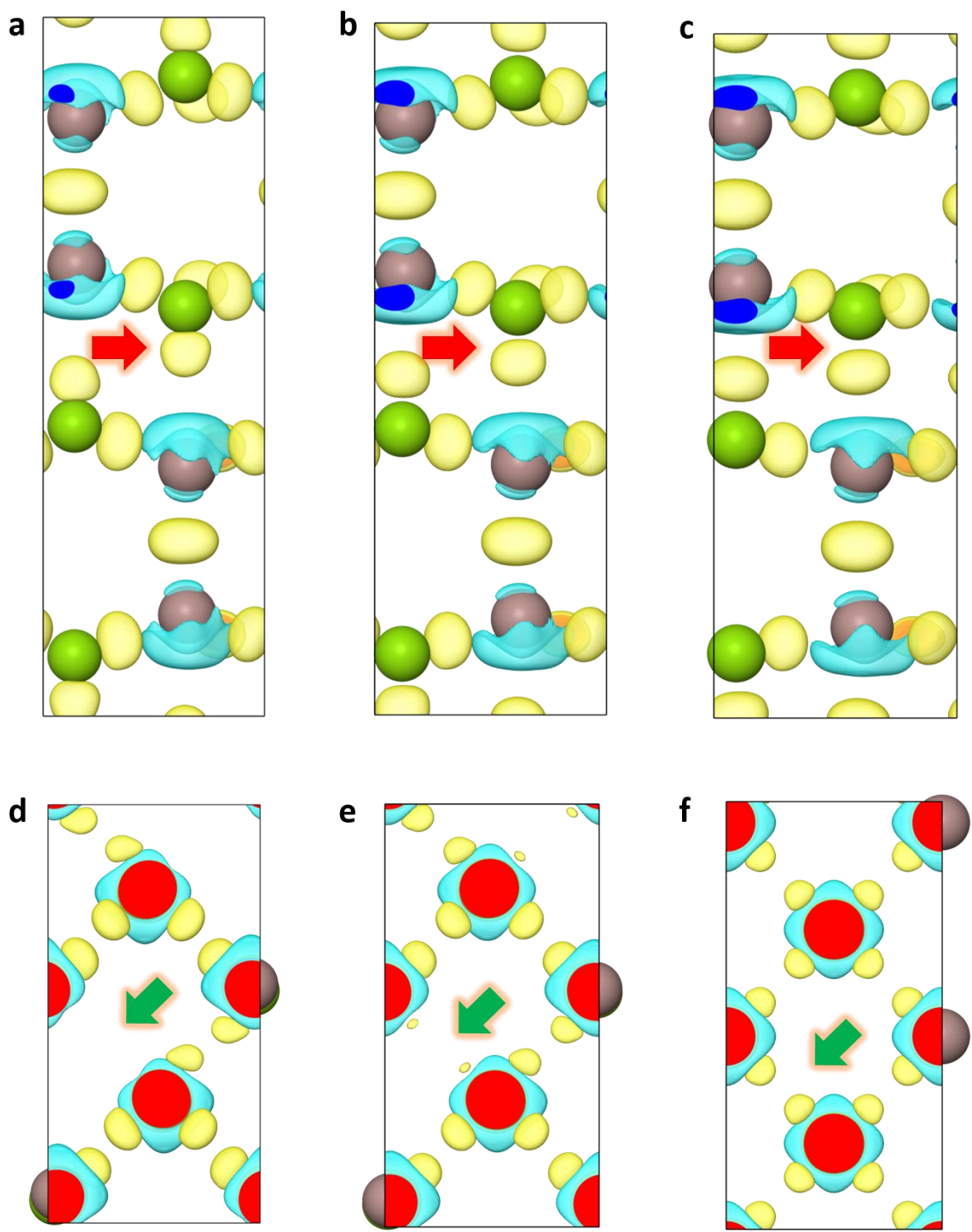

**Fig. S7| Differential charge density for selected configurations in the MEP for OR-2**. **a-c**, Differential charge density of the configurations before the saddle point, at the saddle point, and after the saddle point along the MEP. **d-f**, Differential charge density of the configurations for the last three configurations along the MEP.